\documentclass[journal]{IEEEtran}
\normalsize
\usepackage{cite,comment}
\usepackage{tikz,graphicx,subcaption}
\usetikzlibrary{arrows, positioning}
\usepackage[cmex10]{amsmath}
\usepackage{amssymb, amsthm}
\usepackage{url}
\def\nbb{{\mathbf{b}}}
\def\nbc{{\mathbf{c}}}

\def\nbm{{\mathbf{m}}}

\def\nbx{{\mathbf{x}}}
\def\nby{{\mathbf{y}}}
\def\nbz{{\mathbf{z}}}
\def\nb0{{\mathbf{0}}}
\def\nb1{{\mathbf{1}}}

\def\nbI{{\mathbf{I}}}

\def\nbW{{\mathbf{W}}}

\def\ncalC{{\mathcal{C}}}

\def\ncalN{{\mathcal{N}}}

\def\nbbC{{\mathbb{C}}}

\def\nbbP{{\mathbb{P}}}

\newtheorem{ndef}{Definition}

\newtheorem{eg}{Example}

\begin{document}

\title{Channel Coding meets Sequence Design via Machine Learning for Integrated Sensing and Communications}
\author{Sundar~Aditya,~\IEEEmembership{Member,~IEEE,}
        Morteza~Varasteh,~\IEEEmembership{Member,~IEEE,}
        and~Bruno~Clerckx,~\IEEEmembership{Fellow,~IEEE}%
\thanks{This work was funded by UKRI grants EP/X040569/1, EP/Y037197/1, EP/X04047X/1 and EP/Y037243/1.}%
\thanks{S. Aditya and B. Clerckx are with the Dept.~of Electrical and Electronic Engineering, Imperial College London, London SW7 2AZ, U.K. (e-mail: \{s.aditya, b.clerckx\}@imperial.ac.uk).}%
\thanks{M. Varasteh is with the School of Computer Science and Electronic Engineering, University of Essex, Colchester CO4 3SQ, U.K. (email: m.varasteh@essex.ac.uk)}}

\maketitle

\begin{abstract}
For integrated sensing and communications, an intriguing question is whether information-bearing channel-coded signals can be reused for sensing -- specifically ranging. This question forces the hitherto non-overlapping fields of channel coding (communications) and sequence design (sensing) to intersect by motivating the design of error-correcting codes that have good autocorrelation properties. In this letter, we demonstrate how machine learning (ML) is well-suited for designing such codes, especially for short block lengths. As an example, for rate $1/2$ and block length $32$, we show that even an unsophisticated ML code has a bit-error rate performance similar to a Polar code with the same parameters, but with autocorrelation sidelobes 24dB lower. While a length-32 Zadoff-Chu (ZC) sequence has zero autocorrelation sidelobes, there are only 16 such sequences and hence, a $1/2$ code rate cannot be realized by using ZC sequences as codewords. Hence, ML bridges channel coding and sequence design by trading off an ideal autocorrelation function for a large (i.e., rate-dependent) codebook size. 
\end{abstract}

\bstctlcite{IEEEexample:BSTcontrol}

\section{Introduction}
For communications over unreliable (e.g., noise impaired) channels, the transmitted signal is subjected to channel coding in order to accurately retrieve the intended message at the destination. For ranging (i.e., distance estimation), which is a key component of many radar sensing use cases, it is desirable for the transmitted signal to have an \emph{ideal} autocorrelation function (i.e., equal to $\delta[\cdot]$). These requirements have led to the development of two very rich fields, namely,
\begin{itemize}
    \item \emph{Coding Theory}, which has resulted in many families of codes (e.g., Reed-Solomon, Polar etc.) with a range of desirable properties from an error-correcting perspective, such as large minimum distance, polarization etc.; and,

    \item  \emph{Sequence Design}, which has resulted in many families of sequences (e.g., Zadoff-Chu (ZC), Golay complementary etc.) that have ideal or near-ideal autocorrelation function
\end{itemize}
However, in the context of integrated sensing and communications (ISAC), an intriguing question is whether information-bearing channel-coded signals can be reused for ranging, thereby allowing for efficient resource (time, frequency, power) usage. This forces the fields of channel coding and sequence design to intersect by motivating the design of error-correcting codes that have good autocorrelation properties.

Before exploring the design of such codes, we describe the limitations of two obvious approaches for realizing dual-use ISAC signals:
\paragraph{Using good autocorrelation sequences as error-correcting codewords} While this approach can be viewed as a special case of index modulation, a feature common to all families of sequences with good autocorrelation properties is the relative scarcity of sequences for a given length $N$, when compared to the number of possible $N$-length sequences. This is a direct consequence of striving for an ideal autocorrelation function. As an example, for $N = 32$, the number of ZC sequences (which incidentally have an ideal autocorrelation function) equals $16$, which limits the code rate to $(\log_2 16) / 32 = 1/8$. A higher code rate cannot be realized as a one-to-one mapping between messages and ZC sequences cannot be constructed.

\paragraph{Using error-correcting codewords as ranging sequences} Codewords comprising of i.i.d (zero-mean) Gaussian vectors are asymptotically capacity-achieveing (over AWGN channels) at large block lengths. Such codewords also have an asymptotically ideal autocorrelation function at large block lengths due to the law of larger numbers (LLN). However, state-of-the-art error-correcting codes are rooted in finite-field algebra, where the codeword symbols post modulation are non-Gaussian and correlated. In \cite{Adi_ccs_ojcoms_2023}, it was shown that even such codewords have an asymptotically ideal autocorrelation function at large block lengths (i.e., sidelobes decaying as $O(e^{-rN})$ for code rate $r$ and block length $N$), regardless of the code's properties such as its generator check matrix. The sidelobe decay is again governed by LLN. For short block lengths though, LLN is not applicable and the sidelobe levels of state-of-the-art error-correcting codewords in this regime are much higher than what can be achieved through sequence design. Hence, we focus on the short block length regime in this letter.

The above limitations clarify the ideal properties that an ISAC error-correcting code must possess in the short block length regime, namely, 
\begin{itemize}
    \item[1)] good error-correcting capability,
    \item[2)] large codebook size (equal to $2^{rN}$ for arbitary $r,N$), and
    \item[3)] each codeword must have an ideal autocorrelation function.
\end{itemize}
However, using coding theory to design such a code is extremely challenging because for point 1 above, existing finite-field construction techniques ignore the complex-valued modulation (e.g., QAM) used to transmit the codewords. However, for point 3, the complex-valued autocorrelation function of the codewords is modulation-dependent, and the binary-to-complex mapping from message bits to \emph{modulated} codeword is non-linear\footnote{This is true even for linear codes over the binary field.} and intractable. To the best of our knowledge, there is no systematic method to design codes with good error-correction and auto-correlation properties. This, in turn, motivates us to use machine learning to construct such codes. 

\subsection{Related Work}
Machine learning has been widely used in recent years for channel coding (see \cite{Matsumine_Ochiai_ojcomms_2024} for a detailed survey), where it has been especially effective in designing codes over regimes where state-of-the-art model-based techniques are sub-optimal, such as channels with feedback \cite{deepcode}, short block lengths \cite{turboae, productae, pmlr-v139-makkuva21a, hebbar2024deeppolar}. A distinctive feature of using machine learning for channel coding is the generalizability problem, i.e., only a small fraction of all feasible messages are seen in training even for small message lengths (e.g. 32 bits). As a result, it is well-known that achieving block length gain (i.e., bit error rate decreasing with increasing block length at sufficiently high SNR) is extremely challenging. While structural insights from model-based techniques such as turbo principle \cite{turboae}, Plotkin tree construction \cite{pmlr-v139-makkuva21a}, polarization \cite{hebbar2024deeppolar}, etc. have been used to improve the performance of neural codes, block length gain nevertheless remains elusive at high SNR even for small message lengths.

Machine learning has been also used for several niche applications of channel coding, such as joint source-channel coding of text \cite{Farsad_Rao_Goldsmith_2018} and images \cite{Bourt_Kurka_Gunduz_2019, Kurka_Gunduz_jsait_2020, Yang_Bian_Kim_2021, Xu_Ai_2022}, low-delay analog joint source-channel coding \cite{Xuan_Narayanan_2023}, joint communications and power transfer \cite{Varasteh_Hoydis_Clerckx_2020}, and integrated sensing and communications \cite{Kim_etal_2024, bian2025lisac}. 

Our work shares some similarity with \cite{bian2025lisac} in terms of how the loss function is formulated. However, \cite{bian2025lisac} does not consider the autocorrelation properties of the learnt codewords. In particular, it does not address the conflicting requirements of having a large number of codewords, each with very low (if not zero) autocorrelation sidelobes, which is the main contribution of this letter.

\subsection{Notation}
Row vectors and matrices are denoted by lower and upper case bold letters, respectively. Given $\nbx$, (i) $\|\nbx\|$ is its Euclidean norm, and (ii) $\nbx(l)$ is the vector obtained by circularly right-shifting $\nbx$ by $l (\geq 1)$ places. For $\nbx, \nby \in \nbbC^N$, $\langle \nbx, \nby \rangle:= \nbx \nby^H = \sum_{i = 1}^N x_i y_i^*$ denotes their inner product.

\section{System Model}
In general, a code can be parameterized by the tuple $(K,~N, ~g_{\rm enc}(\cdot),~g_{\rm dec}(\cdot))$, where:
\begin{itemize}
    \item[i)] positive integer $K$ denotes the message length (in bits);
    \item[ii)] positive integer $N (\geq K)$ denotes the blocklength (in symbols). The code rate, $r$, equals $K/N$, throughout this letter we consider $r = 1/2$;
    \item[iii)] $g_{\rm enc}(\cdot): \{0,1\}^{K} \rightarrow \nbbC^N$ denotes the encoder mapping that transforms $K$ message bits to an $N$-length (modulated) codeword; 
    \item[iv)] $g_{\rm dec}(\cdot): \nbbC^N \rightarrow \{0,1\}^{K}$ denotes the decoder mapping that attempts to recover the message bits from a noisy version of the codeword.
\end{itemize}
Let $\nbm \in \{0,1\}^{K}$ denote the message vector containing the bits to be encoded, and let $\nbc \in \nbbC^N$ denote the codeword comprising the coded symbols corresponding to $\nbm$. Then, $\nbc = g_{\rm enc}(\nbm)$, and $\hat{\nbm} = g_{\rm dec}(\hat{\nbc})$, where $\hat{\nbm}$ is the estimate of $\nbm$ recovered from $\hat{\nbc}$, a noisy version of $\nbc$. We assume $\hat{\nbc} = \nbc + \nbz$, where $\|\nbc\| = 1$ (unit power constraint) and $\nbz \in \ncalC\ncalN(\mathbf{0}, \sigma^2 \nbI)$. 

\begin{ndef}[Autocorrelation function]
\label{def:acsl}
For a codeword, $\nbc$, its (periodic) auto-correlation function at lag $l \in \{1, \cdots, N-1\}$, denoted by $\chi(l; \nbc)$, is given by:
\begin{IEEEeqnarray}{rCl}
\label{eq:chi}
    \chi(l; \nbc) &:=  \frac{\langle \nbc, \nbc(l) \rangle}{\langle \nbc, \nbc \rangle} 
\end{IEEEeqnarray}
\end{ndef}

\begin{figure*}
    \centering
    \begin{tikzpicture}
        \node [draw,
              align = center,
	           minimum width=1cm,
	           minimum height=0.2cm, 
             ] (enc_ip_linear) at (0,0) {Linear \\ $(Km_s, 2N)$};
        \node [draw,
              align = center,
	           minimum width=1cm,
	           minimum height=0.2cm,
              right = 0.5cm of enc_ip_linear
              ]  (enc_hl1_linear) {Linear \\ $(2N, 2N)$};
        \node [draw,
              align = center,
	           minimum width=1cm,
	           minimum height=0.2cm,
              right = 0.5cm of enc_hl1_linear, 
              ]  (enc_actv1) {$\tanh(\cdot)$};
        \node [draw,
              align = center,
	           minimum width=1cm,
	           minimum height=0.2cm,
              right = 0.5cm of enc_actv1, 
              ]  (enc_hl2_linear) {Linear \\ $(2N, 2N)$};
        \node [draw,
              align = center,
	           minimum width=1cm,
	           minimum height=0.2cm,
              right = 0.5cm of enc_hl2_linear, 
              ]  (enc_actv2) {$\tanh(\cdot)$};
        \node [draw,
              align = center,
	           minimum width=1cm,
	           minimum height=0.2cm,
              right = 0.5cm of enc_actv2, 
              ]  (enc_op_linear) {Linear \\ $(2N, 2N)$};
        \node [draw,
              align = center,
	           minimum width=1cm,
	           minimum height=0.2cm,
              right = 0.5cm of enc_op_linear, 
              ]  (batch_norm) {Batch Norm};
        \node[above left = 0.6cm and 0.1cm of enc_ip_linear.west] (A){};
        \node[above = 0.2cm of enc_ip_linear] {$g_{\rm enc}(\cdot)$};
        \node[below right = 0.6cm and 0.1cm of batch_norm.east] (B){};
        \draw[-] (A) rectangle (B); 
        \draw[>-] (enc_ip_linear.west) -- ++(-0.5,0) node[at end, left]{$\nbm$};
        \draw[->] (enc_ip_linear.east) -- (enc_hl1_linear.west);
        \draw[->] (enc_hl1_linear.east) -- (enc_actv1.west);
        \draw[->] (enc_actv1.east) -- (enc_hl2_linear.west);
        \draw[->] (enc_hl2_linear.east) -- (enc_actv2.west);
        \draw[->] (enc_actv2.east) -- (enc_op_linear.west);
        \draw[->] (enc_op_linear.east) -- (batch_norm.west);
        \draw[->] (batch_norm.east) -- ++(0.5,0) node[at end, right]{$\nbc$};
        \node [draw,
              align = center,
	           minimum width=1cm,
	           minimum height=0.2cm, 
              below = 1cm of enc_ip_linear
             ] (dec_ip_linear) {Linear \\ $(2N, 2N)$};
        \node [draw,
              align = center,
	           minimum width=1cm,
	           minimum height=0.2cm,
              right = 0.5cm of dec_ip_linear
              ]  (dec_hl1_linear) {Linear \\ $(2N, 2N)$};
        \node [draw,
              align = center,
	           minimum width=1cm,
	           minimum height=0.2cm,
              right = 0.5cm of dec_hl1_linear, 
              ]  (dec_actv1) {$\tanh(\cdot)$};
        \node [draw,
              align = center,
	           minimum width=1cm,
	           minimum height=0.2cm,
              right = 0.5cm of dec_actv1, 
              ]  (dec_hl2_linear) {Linear \\ $(2N, 2N)$};
        \node [draw,
              align = center,
	           minimum width=1cm,
	           minimum height=0.2cm,
              right = 0.5cm of dec_hl2_linear, 
              ]  (dec_actv2) {$\tanh(\cdot)$};
        \node [draw,
              align = center,
	           minimum width=1cm,
	           minimum height=0.2cm,
              right = 0.5cm of dec_actv2, 
              ]  (dec_op_linear) {Linear \\ $(2N, 2Km_s)$};
        \node [draw,
              align = center,
	           minimum width=1cm,
	           minimum height=0.2cm,
              right = 0.5cm of dec_op_linear, 
              ]  (softmax) {Softmax};
        \node[above left = 0.6cm and 0.1cm of dec_ip_linear.west] (C){};
        \node[above = 0.2cm of dec_ip_linear] {$g_{\rm dec}(\cdot)$};
        \node[below right = 0.6cm and 0.1cm of softmax.east] (D){};
        \draw[-] (C) rectangle (D); 
        \draw[>-] -- (dec_ip_linear.west) -- ++(-0.5,0) node[at end, left] {$\hat{\nbc}$};
        \draw[->] (dec_ip_linear.east) -- (dec_hl1_linear.west);
        \draw[->] (dec_hl1_linear.east) -- (dec_actv1.west);
        \draw[->] (dec_actv1.east) -- (dec_hl2_linear.west);
        \draw[->] (dec_hl2_linear.east) -- (dec_actv2.west);
        \draw[->] (dec_actv2.east) -- (dec_op_linear.west);
        \draw[->] (dec_op_linear.east) -- (softmax.west);
        \draw[->] (softmax.east) -- ++(0.5,0) node[at end, right]{$\hat{\nbm}$};
    \end{tikzpicture}
    \caption{$g_{\rm enc}(\cdot)$ and $g_{\rm dec}(\cdot)$ form an autoencoder. We consider $K = 16,~ 32$ and $r = K/N = 1/2$.}
    \label{fig:autoencoder}
\end{figure*}
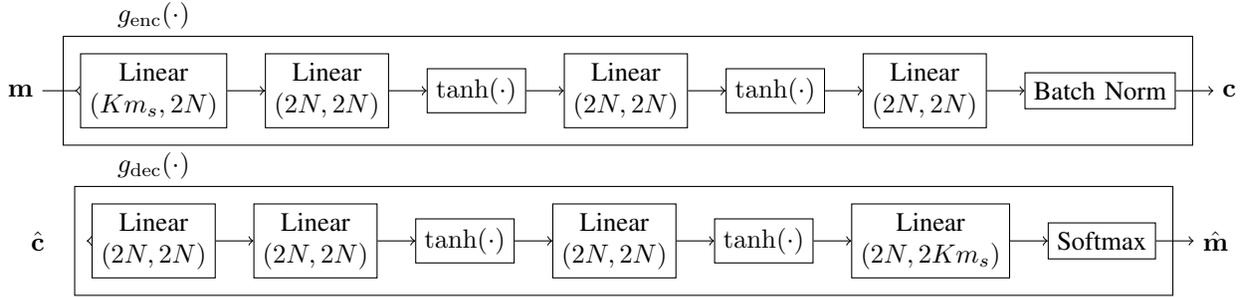

We use neural networks (NN) to construct $g_{\rm enc}(\cdot)$ and $g_{\rm dec}(.)$, as part of an autoencoder. Several NN architectures have been investigated purely from a channel coding perspective, such as feed-forward NNs \cite{pmlr-v139-makkuva21a, hebbar2024deeppolar}, convolutional NNs \cite{turboae}, recurrent NNs \cite{bian2025lisac}, etc. However, to illustrate the potential of NNs for designing codes with good error-correcting and autocorrelation properties, it is sufficient to restrict our attention to (the relatively unsophisticated) feed-forward NNs. Fig.~\ref{fig:autoencoder} illustrates the structure of $g_{\rm enc}(\cdot)$ and $g_{\rm dec}(.)$, whose parameters are optimized to minimize the following loss function:
\begin{align}
\label{eq:loss}
    &L(g_{\rm enc}, g_{\rm dec}) := \lambda \frac{1}{N-1} \sum_{l = 1}^{N-1} |\chi(l;\nbc)|^2 + \frac{(1-\lambda)}{Km_s} \notag \\
    &~\times \sum\limits_{i=1}^{Km_s}[- m_i \log \nbbP(\hat{m}_i=1) - (1-m_i) \log \nbbP(\hat{m}_i=0)],
\end{align}
where the first term (without the $\lambda$ scaling factor) incentivizes codewords with low autocorrelation sidelobe levels (ACSL), and the second term (without the ($1-\lambda$) scaling factor) captures the codewords' error-correcting capability via the cross-entropy between message bit $m_i$ and its decoded estimate $\hat{m}_i$. $\lambda \in [0,1]$, captures the relative importance assigned to the sensing performance of the learned codewords. 

\section{Training}
In each epoch, we alternately optimize $g_{\rm enc}(\cdot)$ and $g_{\rm dec}(\cdot)$ over $N_{\rm enc}$ and $N_{\rm dec}$ iterations, respectively. We consider $K = 16,~32$. For $K = 32$, generalizability over unseen messages is a major challenge as only a small fraction of all possible message vectors ($2^{32}$) are encountered in training. To improve generalizability, we initialize the weights and biases of $g_{\rm enc}(\cdot)$ and $g_{\rm dec}(\cdot)$ to realize a \emph{concatenated code}, defined as follows:

\begin{ndef}[Concatenated Code]
   Let $\nbm = [\nbm_0 ~ \nbm_1]$ denote a length-32 message vector, where $\nbm_0$ and $\nbm_1$ are its left and right halves of length 16. Let $\nbc_0$ and $\nbc_1$ denote the codewords corresponding to $\nbm_0$ and $\nbm_1$, respectively (i.e., for message length 16). Then, the codeword for $\nbm$ defined as $\nbc := [\nbc_0 ~ \nbc_1]$ is referred to as a concatenated code.
\end{ndef}
Clearly, the BER performance of the concatenated code for $K = 32$ matches the BER performance of the code for $K = 16$. Hence, the concatenated code represents a good starting point to seek a final code with better BER performance than $K = 16$ (i.e., block length gain). 

For either $g_{\rm enc}(\cdot)$ or $g_{\rm enc}(\cdot)$, let $\nbW_{32}$ and $\nbb_{32}$ denote the weight and bias corresponding to an arbitrary NN layer for $K = 32$. Similarly, let $\nbW_{16}$ and $\nbb_{16}$ denote the (trained) weight and bias of the corresponding layer for $K = 16$. From Fig.~\ref{fig:autoencoder}, the subscript-32 quantities are twice as large in every dimension as the subscript-16 quantities. Thus, the concatenated code can be realized as follows: 
\begin{align}
    \nbW_{\rm 32} &= \begin{bmatrix}
                     \nbW_{\rm 16} & \mathbf{0} \\
                     \mathbf{0} & \nbW_{\rm 16}
                    \end{bmatrix} \\
    \nbb_{\rm 32} &= \begin{bmatrix}\nbb_{\rm 16} & \nbb_{\rm 16} \end{bmatrix}.
\end{align} 

The training SNR should be low enough to force codewords to be far apart, but not so low that only noise statistics are learnt. Hence, for $K = 16$, we choose a training SNR of $3{\rm dB}$ for the first 200 epochs, which is then increased to ${\rm 6dB}$ for the remaining epochs to improve the BER performance at high SNR. For $K = 32$ however, we maintain a training SNR of ${\rm 3dB}$ throughout to maximize the separation between codewords in our quest for block length gain. A list of hyperparameters is provided in Table~\ref{tab:hyper_param}.
\begin{table*}[h]
    \centering
    \begin{tabular}{|c|l|}
    \hline
      Hyperparameter   & Value \\
    \hline
       Batch size  & 1000 \\
       Optimizer & Adam \\
       No.~of epochs & 400 \\
       ($N_{\rm enc}$, $N_{\rm dec}$) & (10, 50) \\
       Training SNR & \begin{tabular}{ll}
        $K = 16$:   & 3 {\rm dB} for first 200 epochs, 6 {\rm dB} for remaining epochs\\
        $K = 32$:   & 3 {\rm dB} for all epochs 
       \end{tabular}\\
       Learning rate & Initially $10^{-4}$, reducing by a factor of $0.9$ if loss function plateaus over 10 epochs, but going no lower than $10^{-6}$ \\
    \hline
    \end{tabular}
    \caption{List of hyperparameters.}
    \label{tab:hyper_param}
\end{table*}

\section{Results}
For $K = 16,~32$, we compare the error-correcting and autocorrelation properties of the learnt codes with that of Polar codes. For the latter, the codeword bits are interleaved before BPSK modulation (i.e., bit-interleaved coded modulation). The learnt codeword symbols can take arbitrary complex values.

\begin{figure}[h]
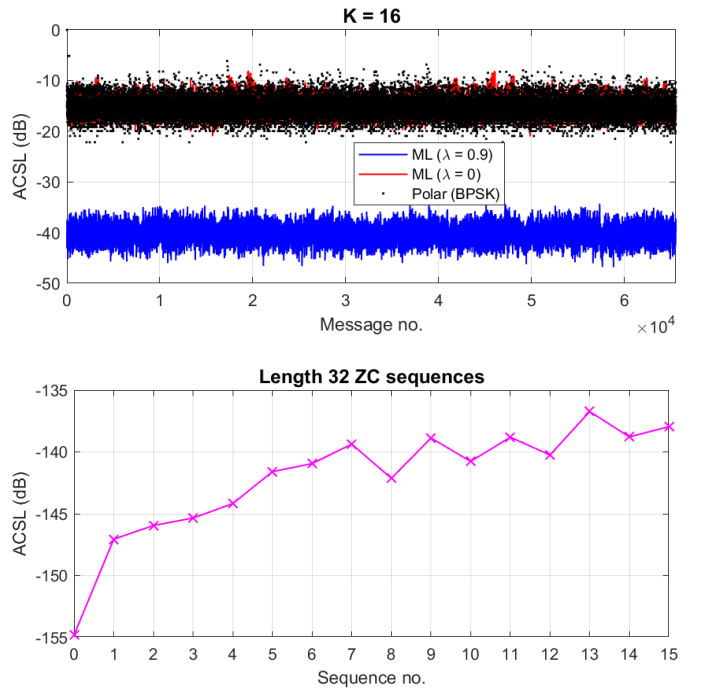

    \centering
    \begin{subfigure}{\linewidth}
        \centering
        \includegraphics[width = \linewidth]{per_cw_acsl_K16_ms1.pdf}
    \end{subfigure}
    \par \bigskip
    \begin{subfigure}{\linewidth}
        \centering
        \includegraphics[width = \linewidth]{ZC_acsl_length32.pdf}
    \end{subfigure}
    \caption{(\textbf{Top}): The ACSL (first term in (\ref{eq:loss})) for each of the $2^{16} = 65536$ codewords of length 32. The median ACSL equals: $-40.08{\rm dB}$ (ML code, $\lambda = 0.9$), $-15.23{\rm dB}$ (ML code, $\lambda = 0$) and $-15.49{\rm dB}$ (Polar code + BPSK). (\textbf{Bottom}): The ACSL of each ZC sequence of length 32. By design, this should be equal to $-\infty {\rm dB}$, the values here are due to limited numerical precision.}
    \label{fig:per_cw_acsl}
\end{figure}

For $K = 16$, Fig.~\ref{fig:per_cw_acsl} (top) plots the ACSL -- i.e., the first term in (\ref{eq:loss}) without the $\lambda$ scaling factor -- of each of the $2^{16} = 65536$ codewords of length 32. The median codeword ACSL for learnt (ML) code with $\lambda = 0.9$ is $-40.08 {\rm dB}$, which is ${24.59\rm dB}$ lower than the corresponding value for a Polar code with BPSK modulation ($-15.49 {\rm dB}$). While the ACSL of a length 32 ZC sequence is much lower (technically $-\infty {\rm dB}$), there are only 16 such sequences (Fig.~\ref{fig:per_cw_acsl}, bottom). Hence, ML helps bridge channel coding and sequence design by trading off an ideal autocorrelation function for a large codebook size (exponential in message length, $K$).

While a larger constellation for the learnt codeword symbols ($\nbbC$ as opposed to $\{-1, 1\}$) provides more options for realizing a near-ideal autocorrelation function, the lower ACSL for the ML code $(\lambda = 0.9)$ is not due to this alone. This is evidenced by the fact that for $\lambda = 0$ (no priority for sensing), the ML codewords -- despite having no constellation restrictions -- have ACSL values very similar to their BPSK modulated Polar counterparts. 

\begin{figure}[h]
    \centering
        \includegraphics[width = \linewidth]{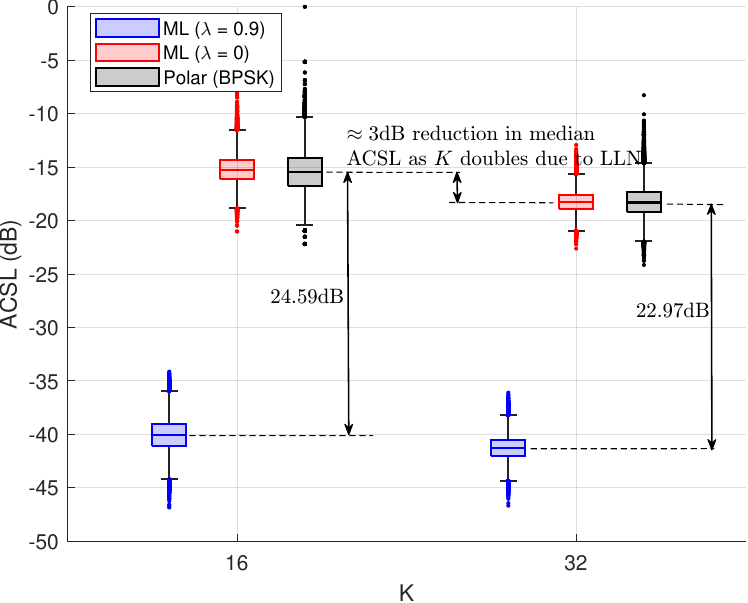}
    \caption{For $K = 16$, the boxplots capture the ACSL distribution of Fig.~\ref{fig:per_cw_acsl}a. For $K = 32$, the boxplots capture the empirical ACSL distribution of codewords corresponding to a common test set of $10^6$ message vectors. The additional reduction in the ACSL for $\lambda = 0.9$ is due to prioritizing sensing in (\ref{eq:loss}).}
    \label{fig:acsl_boxplots}
\end{figure}

For $K = 32$, the codebook size ($2^{32}$) is too large to evaluate the ACSL for each codeword, as in Fig.~\ref{fig:per_cw_acsl}. Hence, in Fig.~\ref{fig:acsl_boxplots}, the boxplots for $K = 32$ capture the empirical ACSL distribution corresponding to a random sample of $10^6$ message vectors, where the \emph{same} set of message vectors were used for the Polar and neural encoders. For Polar codes, the (sample) median ACSL falls by nearly $3{\rm dB}$ as $K$ doubles, consistent with the ACSL decaying as $O(e^{-rN})$ \cite{Adi_ccs_ojcoms_2023}. This trend is mirrored by the ML codes ($\lambda = 0$), as well. Hence, the ACSL corresponding to Polar codes in Fig.~\ref{fig:acsl_boxplots} can be viewed as a baseline achievable through message randomness and LLN alone, with any additional ACSL reduction arising from prioritizing sensing in the first term of (\ref{eq:loss}). Furthermore, the LLN-governed ACSL reduction with increasing block length also highlights the diminishing value of using machine learning to design codes that achieve low ACSL at large block lengths.

\begin{figure}
    \centering
        \includegraphics[width = \linewidth]{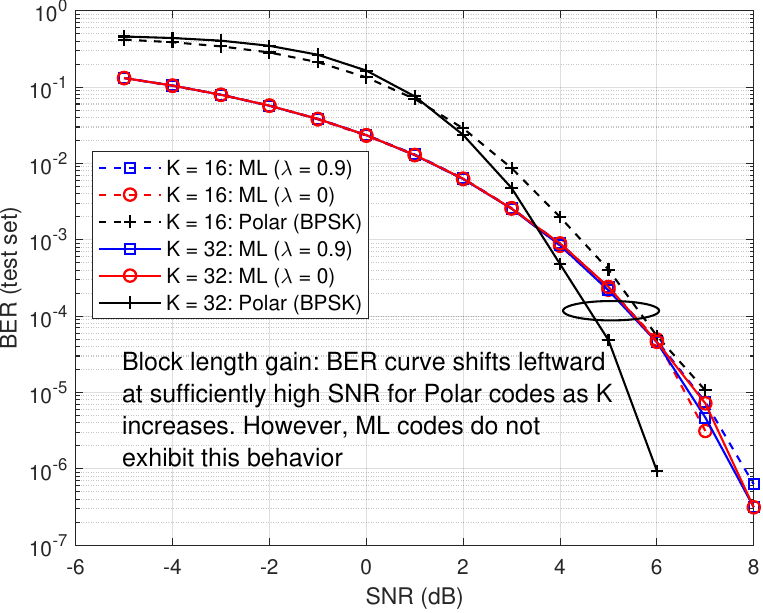}
        \caption{For the ML codes, there is no communications-sensing trade-off -- the BER performance is identical for $\lambda = 0$ and $\lambda = 0.9$. The ML codes outperform Polar codes at low SNR, but do not exhibit block length gain at high SNR.}
    \label{fig:ber}
\end{figure}

Fig.~\ref{fig:ber} compares the BER performance of the learnt and Polar codes over the same test set of $10^6$ message vectors used in Fig.~\ref{fig:acsl_boxplots}. At the outset, we observe that for the learnt codes, there is no loss in BER performance by prioritizing sensing -- i.e., no communications-sening tradeoff. This is due to the fact that the codeword symbols can take arbitrary complex values; hence, there are more options w.r.t simultaneously achieving low ACSL and increasing the pairwise distance between codewords. For $K = 16$, the learnt code generalizes well over unseen SNR by outperforming Polar codes at low SNR and matching Polar code performance at high SNR ($\geq$ 9dB) by having zero errors over the test set. For $K = 32$, the block length gain is clearly seen in Polar codes with the leftward shift of the BER curve relative to $K = 16$ at sufficiently large SNR. However, despite a good starting point (i.e., the concatenated code), the learnt code for $K = 32$ does not exhibit a similar block length gain. 

\begin{figure}
    \centering
    \begin{subfigure}{\linewidth}
        \centering
        \includegraphics[width = \linewidth]{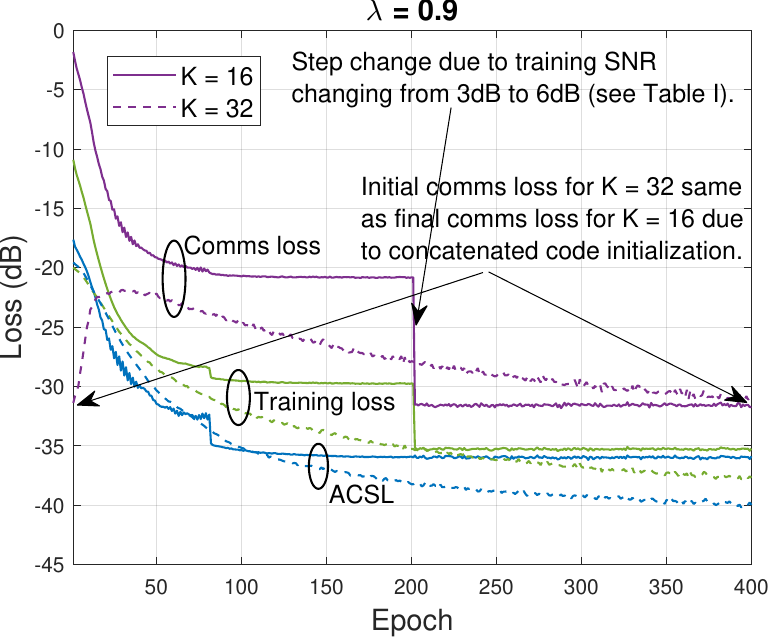}
    \end{subfigure}
    \par \bigskip
    \begin{subfigure}{\linewidth}
        \centering
        \includegraphics[width = \linewidth]{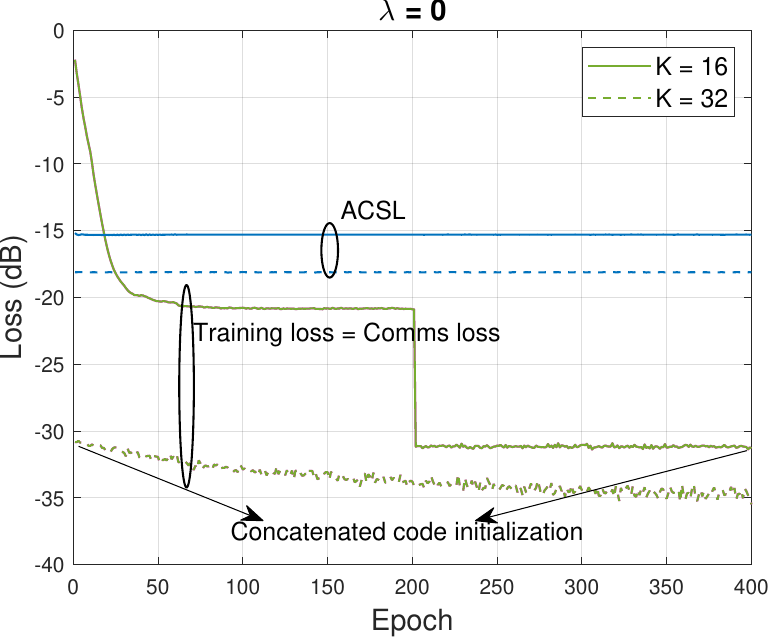}
    \end{subfigure}
    \caption{The evolution of the training loss with concatenated code initialization for $K = 32$.}
    \label{fig:concat}
\end{figure}

Fig.~\ref{fig:concat} illustrates the effect of the concatenated code initialization. For $K = 32$, the initial \emph{communications loss} (i.e., the second term in (\ref{eq:loss}) without the $(1-\lambda)$ scaling factor) at epoch 0 coincides with the final communications loss for $K = 16$, which is consistent with the fact that the concatenated code has the same BER performance as the $K = 16$ code. However, codewords in the concatenated code need not have low ACSL. Hence, for $\lambda = 0.9$, in moving towards the direction of low ACSL (and low training loss), the communications loss initially increases before eventually decreasing. This behavior is not seen for $\lambda = 0$, where the communications loss coincides with the training loss. In both cases, the final communications loss is not very different from the initial value.

On the one hand, the concatenated code initialization guarantees a certain amount of generalizability over unseen messages by producing a code at least as good as the $K = 16$ code in terms of BER performance. However, this initialization seems to be a local minima that is hard to escape. This further underscores the well-documented challenges of achieving block length gain using neural codes. Future works on channel coding for ISAC should focus on achieving block length gain, possibly through a combination of different neural network architectures and curriculum learning along the lines of \cite{hebbar2024deeppolar}.

\section{Summary}
In this letter, we used neural encoders and decoders to design codes with good error-correcting/autocorrelation properties. From a sensing perspective, such neural codes are quite powerful in overcoming the scarcity problem inherent in sequence design -- instead of very few sequences with an ideal auto-correlation function, neural codes produce a far larger number of sequences/codewords (exponential in the message length) with low, but non-zero, ACSL. Crucially, the ACSL of such sequences is far lower than what can be realized through message randomness alone in the short block length regime.

From a communications perspective, it is a well-known challenge to design neural codes that achieve block length gain at high SNR, since most messages are not seen in training even for message lengths as small as 32. This challenge exists even if we require the codewords to have good autocorrelation properties. 

\bibliographystyle{IEEEtran}
\bibliography{IEEEabrv, mmWaveRadCom_bib}
\end{document}